\documentstyle[epsfig,12pt]{article}

\begin{document}
\centerline{ \LARGE \bf Cosmology with a time-varying speed of light
\footnote{To appear in the proceedings of COSMO98, D. Caldwell Ed.}
}

\begin{center}
{Andreas Albrecht}\\
Department of Physics\\
One Shields Ave.\\
University of California\\
Davis CA 95616
\end{center}


\begin{abstract}
Cosmic inflation is the only known mechanism with the potential to
explain the very special initial conditions which are required at the
early stages of the evolution of our universe. This article
outlines my work with Joao Magueijo which attempts to construct an
alternative mechanism based on a time-varying speed of light.
\end{abstract}

\section*{Introduction}
As we reconstruct the past history of the universe, we learn that the
universe as we see it today must have evolved from a special state
that was extremely flat and homogeneous.  Such a state is highly
unstable toward both the formation of inhomogeneities and evolution away
from flatness, because the influence of gravity tends to drive matter
away from a flat homogeneous state.  Until the advent of inflation,
cosmologists had no way of explaining how the universe could have
started out so precisely balanced in a state of such high instability.
 
Inflation addresses this issue by changing the story at very early
times.  During an early period of inflation, the matter is placed in a
peculiar potential dominated state where the effects of gravity are
very different.  During inflation, flatness is an attractor, and the
deviations (or perturbations) from homogeneity destined for our
observable universe can be calculated.  With a suitable tuning of
model parameters, the perturbations can be given a sufficiently small
amplitude and even naturally acquire a spectrum which gives good
agreement with observations.

The theory of cosmic perturbations has for some time now benefited 
from the existence of alternative models. The presence of alternatives
has made it possible to systematically evaluate each alternative, and
has even helped us to discover the most fundamental way in which inflationary
theory could be falsified\cite{aadi}. So far, things are looking very
good for an inflationary origin for the cosmic perturbations.

But inflation offers us much more than a theory of perturbations.  It
also is supposed to explain the origin of the flatness and overall
smoothness of the universe.  In this role, inflation theory has faced
no serious competition.  While this fact in itself might be seen as a
success, it would certainly be much more gratifying if the significant
place in the theoretical landscape currently occupied by inflation
could be earned by doing better than some serious contenders.  After
all, inflation is just the first idea we have had to explain these
cosmic puzzles.

The above reasoning has motivated me to search for alternatives for
some time now.  So far everything that I have tried has ended up being
just another version of standard inflation, once it was forced into a
workable form.  This experience encourages the view that
inflation might be {\em the} unique mechanism by which the initial condition
problems can be addressed, and also explains the extreme nature of the
idea I outline below.

\begin{figure}[b!] 
\centerline{\epsfig{file=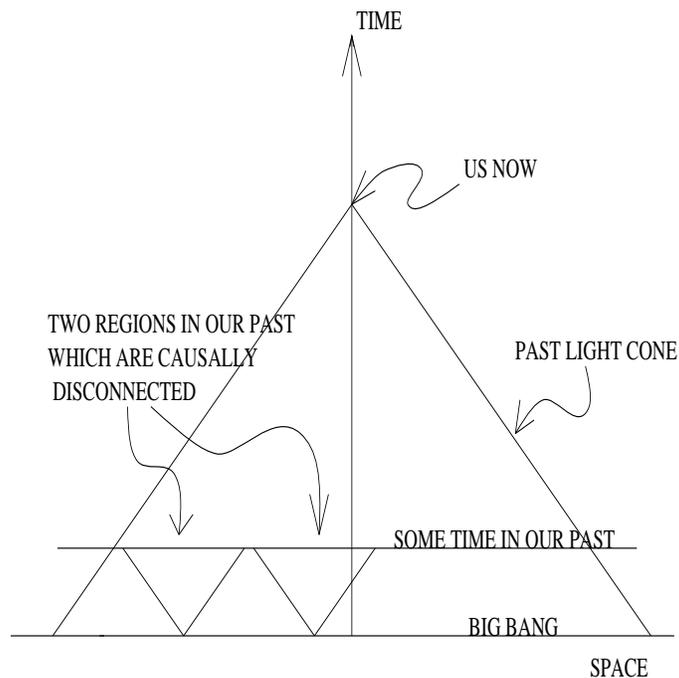,height=3.5in,width=3.5in,angle=-90}}
\vspace{10pt}
\caption{The horizon problem in standard cosmology.}
\label{fig1}
\end{figure}

\begin{figure}[b!] 
\centerline{\epsfig{file=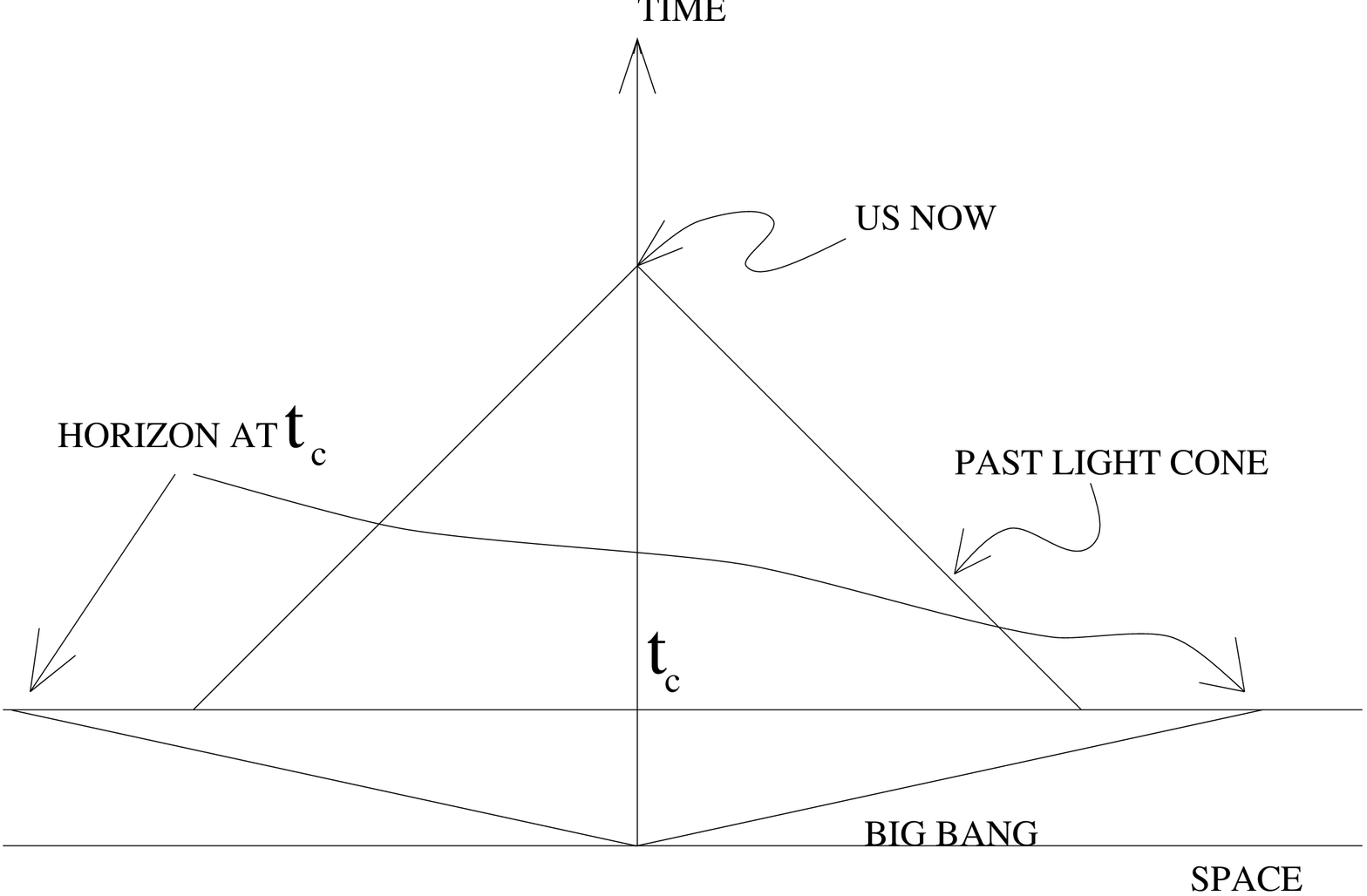,height=3.5in,width=3.5in,angle=-90}}
\vspace{10pt}
\caption{The horizon problem solved by a time-varying speed of light.}
\label{fig2}
\end{figure}

The idea I present here starts with a very simple observation. A
statement of the unusual nature of the initial conditions in standard
cosmology usually carries with it a description of the ``horizon
problem''.  Basically, in the standard big bang model any mechanism
which operates in the early universe and attempts to ``set up'' the
correct initial conditions would have to act acausally, because what
we observe today is composed of many causally disconnected regions in
the early universe (see Figure \ref{fig1}).  Inflation solves this
problem because a period of 
superluminal expansion radically changes the causality structure of the
universe.  Another way of changing the causality structure is to have
light travel faster in the early universe (Figure \ref{fig2}).


Joao Magueijo and I have pursued this idea, to the extent of setting
up a phenomenological model of how physics might look with a time
varying speed of light (VSL)\cite{am}.  Interestingly, we have found that our
model exhibits energy non-conservation of just the sort that can fill in
energy deficits and pull down energy peaks to produce a flat
homogeneous universe from a wide range of initial conditions. 

Of course our model also breaks Lorentz invariance, which may seem
unreasonable to may physicists.  In defense of choosing this radical
direction, let me comment that many theorists are prejudiced {\em
against} the idea that the spacetime continuum is really a continuum
down to arbitrarily small scales.  Any deviation from a true continuum
would necessarily break Lorentz invariance.  In particular many ideas
about our 3+1 dimensional world that are coming from superstring
theory (and its offspring) suggest that properties of this (3+1D) manifold
are emergent as a low-energy limit of something quite different.  If
the VSL picture really takes root, the picture I describe below could
well be a phenomenology of the dynamics of the universe as the 3+1
manifold we inhabit emerges from very different physics governing high
energies.  In this picture the speed of light varies in the early
universe and then holds constant, so that the standard cosmology can
proceed.  This much is in the same spirit as inflation, where a period
of unusual physics is placed in the early universe, without changing
the standard physical picture which does an excellent job of
explaining many aspects of the universe.

I should mention that the idea of using a varying speed of light to
explain initial conditions first appears in print in a paper by
Moffat\cite{mof}.  His paper takes the idea in a
somewhat different direction than we have.  Also, subsequent work by
Barrow and Magueijo has taken VSL in a variety of different
directions.  I will mention this briefly in the final section.

\section*{Our prescription}
To pursue the idea of VSL, Magueijo and I have used the following
simple prescription:  VSL models necessarily have preferred frame,
because Lorentz invariance is broken.  We assume that in that special
frame, the Lagrangian is the same as usual, with the substitution $c
\rightarrow c(t)$.  We also assume that the dynamics of $c$ do not
affect the curvature, so that the Riemann tensor and the Ricci scalar
are to be computed (in the preferred frame) with $c$ held fixed.

This scheme is spelled out in \cite{am}.  There we carefully
discuss the question of why this scheme is {\em not} simply ordinary
physics under a strange reparameterization, and thus why what I
describe below can not be viewed as simply an unusual way of
describing inflation. 

\section*{Cosmological Equations}

Under our VSL scheme energy is not conserved when $c$ is varying.  In
a cosmology which is Robertson-Walker in the preferred frame we get 

\begin{equation}\label{cons1}
\dot\rho+3{\dot a\over a}{\left(\rho+{p\over c^2}\right)}=
-\rho{\dot G\over G}+{3Kc^2\over 4\pi G a^2}{\dot c\over c}.
\end{equation}

This equation is the usual equation for energy
conservation when $c$ and $G$ are constant, and I have included the
term in $\dot G$ for future reference.  To observe the effect on the
flatness of the universe, we look at the evolution of $\epsilon \equiv
\Omega - 1$ ($\Omega \equiv \rho/\rho_c$):

\begin{equation}\label{epsiloneq}
\dot\epsilon=(1+\epsilon)\epsilon {\dot a\over a} 
{\left(1+3w\right)}+2{\dot c\over c}\epsilon
\label{edot}
\end{equation}
where we have taken $p=w\rho c^2$ with constant $w$.  Here we can see
how in standard big bang cosmology ($w > -1/3$, $\dot c = 0$)
$\epsilon = 0$ is an unstable fixed point, leading to the need to tune
$\epsilon$ to extremely small values at the beginning, in order to
match a value of $\epsilon$ which is not large even today.  Eqn. \ref{edot}
also shows how inflation makes $\epsilon = 0$ an attractor, and how
$\dot G$ drops out of the equation, making at least this version of a
varying $G$ ineffective at producing the desired effect.  One can also
see how negative values of $\dot c/ c$ will make $\epsilon = 0$
an attractor.

There is also an interesting effect on the cosmological constant.  In
order to discuss this, we must be careful about which constant we are
talking about.  
\begin{equation}
S=\int dx^4 \sqrt{-g}{\left( {c^4 (R+2\Lambda_1)\over 16\pi G}
 +{\cal L}_M + {\cal L}_{\Lambda_2}\right)}
\label{act}
\end{equation}
Equation \ref{act} shows the action in the preferred frame,
where ${\cal L}_M$ is the matter fields Lagrangian. 
The term in $\Lambda_1$ is a geometrical cosmological constant,
as first introduced by Einstein. The term in $\Lambda_2$ represents
the vacuum energy density of the quantum fields \cite{steve}.  VSL is
only able to affect $\Lambda_1$, and we simply call this $\Lambda$ in
what follows.
If we define $\epsilon_\Lambda=\rho_\Lambda/\rho_m$ where
$\rho_\Lambda={\Lambda c^2\over 8\pi G}$
we find
\begin{equation}\label{epslab}
\dot \epsilon_\Lambda =\epsilon_\Lambda{\left(
3{\dot a\over a}(1+w)+2{\dot c\over c}{1+\epsilon_\Lambda
\over 1+\epsilon}\right)}.
\end{equation} 
Ordinary cosmology has a $\Lambda$ problem in the sense that $\Lambda$ rapidly
comes to dominate, and must be tuned initially in order not to be
super-dominant today. Here we can see that inflation, with $-1 \leq w
\leq -1/3$, cannot provide the necessary tuning, while VSL
($\dot c < 0$) can have that effect.  This, by the way also helps
illustrate how VSL is not physically equivalent to inflation.

We have also considered the evolution of perturbations.  We have found
that for a sudden change in $c$, the density contrast $\Delta$ obeys
\begin{equation}
{\Delta^+\over\Delta^-}={c^+\over c^-}.
\end{equation}
For the large variation in $c$ required to produce a flat universe, we
have found that the universe has all perturbations reduced to
unobservable levels.  This leave a blank slate which requires
something like the defect models of cosmic structure formation to
provide perturbations\cite{vs,bra}.

\section*{Scenario Building}

So far I have discussed the machinery of VSL, but how can this be
turned into a cosmological scenario?  One can start the discussion by
considering a sudden transition between $c^{-}$ and $c^{+}$.  Let us
assume for a moment that before the transition we have a flat FRW
universe.  We find under these circumstances the temperature obeys
\begin{equation}
T^+/T^-= c^{-}/c^{+}.
\end{equation}
If we want $T^+ \approx T_{Planck}$ and require $c^{-}/c^{+} > 10^{60}$
to solve the flatness problem, on is starting with a very cold
$T^{-}$.  Immediately, one can see that fine tuning is required to have
a cold flat universe before the transition, and this scenario does not
make much sense.  Interestingly, unlike inflation Equation \ref{edot}
shows that VSL can create energy even in a empty open ($\epsilon =
-1$) universe.  It might be more interesting to build scenarios based
on that starting point.  Just as inflation has seen the
scenario-building change radially over the years, we feel there is a
lot to be learned about how to implement VSL before we understand the
best scheme to use.  What we have so far is a very interesting
mechanism that can move the universe toward a flat homogeneous state.

\section*{Discussion and Conclusions}

Since our paper there have been a number of other
publications on VSL. One new direction pursued by Barrow and
Magueijo\cite{b1,bm1} looks at a possible power-law $c(t)$ that could
have $c$ varying even today.  They have found interesting attractor
solutions which keeps $\Omega_{\Lambda}$ at a constant fraction of the
total $\Omega$, but they have their work cut out for them understanding
primordial nucleosynthesis in that model.
Also, Moffat has further explored the idea of spontaneous breaking of
Lorentz symmetry\cite{m2}.

Probably the greatest problem with the VSL idea is that we have no
fundamental picture of what makes $c$ vary.  The phenomenological
treatment in \cite{am} does not address this question.  What we can
say is our work shows that there are interesting cosmological rewards
for considering a time varying speed of light, and with that
motivation it may be possible to make some interesting discoveries.

Despite this problem, it is already possible to falsify at least the
fast-transition version of VSL.  Since the perturbations turn out
to be infinitesimal after the transition, structure formation must be left
to active models which have their own characteristic signatures that
differentiate them from inflation\cite{acfm}.  If the microwave
background comes out with characteristic inflationary features, these
VSL models will be ruled out.

This work was supported in part by UC Davis.  I would like to thank
Joao Magueijo for a fruitful collaboration, and Richard Garavuso for
his comments on this manuscript.


\begin{thebibliography}{99}
\bibitem{aadi}A. Albrecht, How to falsify scenarios with primordial
fluctuations from inflation,
in Critical dialogues in cosmology, ed. Neil Turok, PUP Princeton.
\bibitem{am}A. Albrecht and J. Magueijo, PRD 59, 43516 (1999)
\bibitem{mof}J. Moffat, International Journal of Physics D, Vol. 2, No. 3
  (1993) 351-365; Foundations of Physics,
  Vol. 23 (1993) 411. 
\bibitem{steve}S. Weinberg, Theories of the cosmological constant,
in Critical dialogues in cosmology, ed. Neil Turok, PUP Princeton.
\bibitem{vs}A. Vilenkin and E.P.S. Shellard, Topological defects
in Cosmology, CUP, (1996).
\bibitem{bra}R.A. Battye, J.Robinson, A. Albrecht
Phys.Rev.Lett. 80 (1998) 4847-4850
\bibitem{b1}J. D. Barrow PRD 59, 43515 (1999)
\bibitem{bm1} J. D. Barrow, J. Magueijo ``Solving the Flatness and
Quasi-flatness Problems in Brans-Dicke Cosmologies with a Varying
Light Speed''  astro-ph/9901049,
Phys.Lett. B447 (1999) 246, Phys. Lett. B443 (1998) 104
\bibitem{m2} J. W. Moffat ``Varying Light Velocity as a Solution to
the Problems in Cosmology'' 
astro-ph/9811390
\bibitem{acfm}A. Albrecht, D. Coulson, P. Ferreira, J. Magueijo
Phys.Rev.Lett. 76 (1996) 1413-1416
\end{thebibliography}
\end{document}